\documentclass[prd,twocolumn,showpacs,preprintnumbers,amsmath,amssymb,nofootinbib]{revtex4}
\usepackage{graphicx}
\usepackage{dcolumn}
\usepackage{bm}

\newcommand{\units}[1]{\ensuremath{\,\mathrm{#1}}}

\begin{document}


\title{Nucleon mass:\\from lattice QCD to the chiral limit}

\author{M. Procura}
 \email{Massimiliano.Procura@ph.tum.de}
\author{B. U. Musch}
 \email{Bernhard.Musch@ph.tum.de}
\author{T. Wollenweber}
\author{T. R. Hemmert}
 \email{Thomas.Hemmert@ph.tum.de}
\author{W. Weise} 
 \email{Wolfram.Weise@ph.tum.de}

\affiliation{
	Physik-Department, Theoretische Physik \\
	Technische Universit\"{a}t M\"{u}nchen, D-85747 Garching, Germany
	}
\date{\today}

\begin{abstract}
Previous extrapolations of lattice QCD results for the nucleon mass to the physi\-cally relevant region of small quark masses, using chiral effective field theory, are extended and expanded in several directions. A detailed error analysis is performed. An approach with explicit $\Delta(1232)$ degrees of freedom is compared to a calculation with only pion and nucleon degrees of freedom. The role of the $\Delta(1232)$ for the low-energy constants of the latter theory is elucidated. The consistency with the chiral perturbation theory analysis of pion-nucleon scattering data is examined. 
It is demonstrated that this consistency can indeed be achieved if the $\Delta(1232)$ dominance of the P-wave pion-nucleon low-energy constant $c_3$ is accounted for. Introduction of the $\Delta(1232)$ as an explicit propagating degree of freedom is not crucial in order to describe the quark mass dependence of the nucleon mass, in contrast to the situation with spin observables of the nucleon.
The dependence on finite lattice volume is shown to yield valuable additional constraints. What emerges is a consistent and stable extrapolation scheme for pion masses below 0.6 GeV. 
\end{abstract}
	

\maketitle

\section{Introduction}


In recent years the nucleon mass has been in the focus of steadily improving lattice QCD computations. Technical limitations have so far restricted lattice QCD results to quark masses larger than five times the physically relevant masses of the light quarks, $m_{u,d} < 10\units{MeV}$. Under such conditions, the nucleon masses produced on the lattice
are considerably larger  ($M_N > 1.2\units{GeV}$) than the physical one. Systematic extrapolations guided by well-defined rules of low-energy QCD are necessary in order to bridge this gap. Several versions of such extrapolations have been developed in recent years \cite{PHW,BHM,LTY}. They differ partly in details but agree on the basics, namely on the relevance of one-loop chiral pion-nucleon dynamics in determining the dependence of $M_N$ on the pion mass $m_\pi$. 

The connection between $m_\pi$ and the $u$- and $d$-quark masses (for which we take their average,
$m_q = (m_u + m_d)/2$), is given in leading order by PCAC and the Gell-Mann--Oakes--Renner relation,
$m_\pi^2 f_\pi^2 = -2 m_q \langle\bar{q}q\rangle$, with the pion decay constant $f_\pi$ and the chiral condensate $\langle\bar{q}q\rangle$ both taken at the chiral limit. Accurate lattice QCD results for the pion mass as a function of the quark mass $m_q$ \cite{Lusch,Gock,Aoki} demonstrate that this leading linear relation between $m_\pi^2$ and $m_q$ remains remarkably stable even at large quark masses. Corresponding lattice data are consistent with one-loop chiral perturbation theory at next-to-leading order, up to $m_\pi \sim 0.5 \units{GeV}$ \cite{Lusch,Bernhard}. Moreover, the data continue to display the leading-order (linear) PCAC behavior even  beyond this margin, for reasons not yet understood in detail. We will therefore accept this as a fact when we later compare our theoretical $M_N(m_\pi)$ with lattice results.

The present work updates and extends previous investigations \cite{PHW,QCDSF} in several respects:
\begin{itemize}
\item{The overall consistency of parameters and low-energy constants in the expansion of $M_N(m_\pi)$ with those extracted from pion-nucleon scattering, is carefully examined.}
\item{The role of explicit $\Delta(1232)$ degrees of freedom in the one-loop nucleon self-energy is studied.}
\item{A systematic error analysis for the extrapolation from the lattice data through the physical point
down to the chiral limit is performed.}
\item{A study of finite volume effects in comparison with lattice data provides interesting additional constraints.} 
\end{itemize} 
Issues of convergence when carrying chiral expansions over relatively large ranges of quark masses, will also be addressed.

The following section briefly summarizes the framework of the present approach, covariant baryon chiral perturbation theory (B$\chi$PT) using infrared regularization. We omit derivations of basic equations which have already been reported
elsewhere \cite{PHW}. Section \ref{sec-erranalysis} describes the detailed error analysis
of the extrapolation from lattice data down to the physical region. Section \ref{sec-empconstr} discusses connections with the analysis of low-energy pion-nucleon scattering. Section \ref{deltasec} introduces the $\Delta(1232)$ as an explicit degree of freedom and examines its relevance to the chiral extrapolation
of the nucleon mass. Section \ref{sec-voldep} investigates effects induced by varying the finite volume of the lattice, and section \ref{sec-voldep2} completes the analysis by exploring the effects of avoiding the inclusion of the physical nucleon mass as input. Conclusions are drawn in section \ref{sec-conclusions}.

\section{The nucleon mass in chiral perturbation theory}
\label{sec-nucmass}

In previous work \cite{PHW} we have investigated the quark mass dependence of the nucleon mass using chiral effective field theory in the two-flavor, pion-nucleon sector. A fit has been performed to lattice data selected according to the largest available lattice volumes and smallest accessible lattice spacings, in order to be as close as possible to the infinite volume and continuum limits for which these $\chi$PT calculations are designed. We first present a brief summary of this approach.

In ref.\,\cite{PHW} a one-loop calculation has been performed, taking into account diagrams up to chiral order $p^4$ in covariant B$\chi$PT with infrared regularization \cite{BL}. The following result is obtained for $M_N(m_\pi)$ when expanded in powers of $m_\pi$:
\begin{align}
	M_N^{(4)}
	\;=\;& M_0-4c_1 m_\pi^2-\frac{3 g_A^2}{32 \pi f_\pi^2 }m_\pi^3\nonumber\\
	&+ \Big[4\, e_1^{(4)}(\lambda)-\frac{3}{64\pi^2 f_\pi^2}\left(\frac{g_A^2}{M_0}-\frac{c_2}{2}\right)  \nonumber \\
	&- \frac{3}{32 \pi^2 f_\pi^2 }\left(\frac{g_A^2}{M_0} - 8 c_1+c_2+4c_3\right)\ln{\frac{m_\pi}{\lambda}}
		\Big]\,m_\pi^4 \nonumber\\
	&+ \frac{3 g_A^2}{256 \pi f_\pi^2 M_0^2} m_\pi^5 + {\cal{O}}(m_\pi^6)~~.
	\label{massp4}
\end{align}
This formula requires the following input:
\renewcommand{\labelenumi}{\roman{enumi})}
\begin{enumerate}
	\item the nucleon axial vector coupling constant $g_A$ and the pion decay constant $f_\pi$, both taken in the chiral $(m_\pi = 0)$ limit. In practice we use their physical values, $g_A = 1.267$ and $f_\pi = 92.4\units{MeV}$, for guidance and examine variations around these values;
	\item the nucleon mass in the chiral limit, $M_0$;
	\item the low-energy constants $c_1$, $c_2$ and $c_3$. Here $c_1$ is closely linked to the pion-nucleon sigma term, while $c_{2,3}$ encode information primarily on the $\Delta(1232)$ excitation in P-wave pion-nucleon scattering;
	\item a parameter $e_1^{(4)}(\lambda)$, combining three different couplings of the most general ${\cal O}(p^4)$ B$\chi$PT Lagrangian (see ref.\,\cite{PHW}). $e_1^{(4)}(\lambda)$ represents unresolved short distance dynamics. This term compensates the logarithmic dependence on the renormalization scale $\lambda$ so as to ensure scale independence of the result.
	\end{enumerate}

The present work builds upon a successful fit of eq.\,(\ref{massp4}) to unquenched two-flavor lattice results, referred to as ``fit II'' in ref.\,\cite{PHW}. The input lattice data with improved Wilson fermions (points 19, 41, 8 and 23 in ref.\,\cite{QCDSF} and table~\ref{tab-latticedat}) are chosen according to the following criteria: small lattice spacing, $a<0.15\units{fm}$, and large spatial lattice size, $m_\pi L>5$. We have considered only the smallest available pion masses, $m_\pi < 0.6\units{GeV}$.
In order to avoid an under-determined fit we fix some parameters. In particular,
$c_1$ is eliminated by substituting the empirical nucleon mass into eq.\,(\ref{massp4}) at the physical value of the pion mass,
$c_2$ is set equal to $3.2\units{GeV^{-1}}$ as determined in \cite{FMS},
and $c_3$ is fixed at $-3.4\units{GeV^{-1}}$, according to the $NN$ phase shift analysis in ref.\,\cite{EM}.
Employing the physical values $f_\pi = 92.4\units{MeV}$, $g_A = 1.267$, only two parameters
remain free:  $M_0$ and $e_1^{(4)}(\lambda)$ (we chose a renormalization scale $\lambda=1\units{GeV}$ throughout this work). 

In eq.\,(\ref{massp4}) we have truncated the ${\cal O}(p^4)$ infrared regularized expression at $m_\pi^5$ in order to avoid further unconstrained counter terms, cf. ref.\,\cite{PHW}. We note that the truncation of $M_N^{(4)}$ at $\mathcal{O}(m_\pi^4)$ coincides with the $\mathcal{O}(p^4)$ result in heavy baryon chiral perturbation theory (HB$\chi$PT), see ref.\,\cite{BHM}. We stress that in the context of baryon $\chi$PT with infrared regularization, recoil corrections are part of the same order in the corresponding perturbative, diagrammatic expansion. Manifestly covariant and non-relativistic methods just lead to a different organization of the perturbation theory. We have checked that fits based on the HB$\chi$PT ${\cal O}(p^4)$ expression give results that are compatible with the analysis presented here. Thus the term proportional to $m_\pi^5$ in eq.\,(\ref{massp4}), {\it i.e.} the leading recoil correction to the non-relativistic result, plays no significant numerical role.

The application of chiral perturbation theory to interpolations spanning a rather large interval of quark masses has always been a point of concern. 
In the next section we examine the degree of convergence as the calculation of the nucleon mass evolves order by order in the chiral expansion. In preparation of this study we first recall further basic results discussed in ref.\,\cite{PHW}.
At chiral order $p^2$, we have 
\begin{equation}
	M_N^{(2)} = M_0-4c_1m_\pi^2~
	\label{massp2} 
	\end{equation}
and the one-loop expression at ${\cal O}(p^3)$ using infrared regularization, expanded in powers of $m_\pi$, is

\begin{align}
	M_N^{(3)} & = M_0-4c_1 m_\pi^2-\frac{3 g_A^2}{32 \pi f_\pi^2 }m_\pi^3 \nonumber\\
	& + \left[4\,e_1^{(3)}(\lambda)-\frac{3 g_A^2}{64\pi^2 f_\pi^2 M_0}\left(1+ 2\ln{\frac{m_\pi}{\lambda}}\right)
		\right]\,m_\pi^4 \nonumber\\
	& + \frac{3 g_A^2}{256 \pi f_\pi^2 M_0^2} m_\pi^5 + {\cal{O}}(m_\pi^6)~~.
	\label{massp3} 
\end{align} 

Here, $e_1^{(3)}(\lambda)$ absorbs the $\lambda$-dependence at $\mathcal{O}(m_\pi^4)$, which differs from that of $e_1^{(4)}(\lambda)$ in eq.\,(\ref{massp4}).

\section{Error analysis}
\label{sec-erranalysis}

In this section a systematic error analysis is performed in order to test the reliability of the interpolation between lattice results and the physical nucleon mass based on eq.\,(\ref{massp4}).

We are confronted with uncertainties from two different sources of errors, namely {\it input uncertainties} and {\it theoretical uncertainties}, which need to be distinguished throughout the analysis because they require different treatments. Input uncertainties result from statistical errors of lattice calculations\footnote{At the moment we cannot estimate the intrinsic systematic uncertainties in the lattice calculations.} and from limited empirical information on low-energy constants. Theoretical uncertainties come from two sources: First, they arise because our fit function derives from a perturbative expansion and is therefore subject to corrections of higher order. Secondly, we are dealing with an effective field theory, which has a limited range of applicability.
Our analysis is performed in two steps. 
First, we compare different orders in the perturbative expansion of $M_N(m_\pi)$ focusing only on pion masses up to $0.6\units{GeV}$. This range is consistent with the conclusions drawn in ref.\,\cite{BHM}. In a second step we treat input uncertainties.

\begin{figure}[tb]
	\centering
	\includegraphics[width=\columnwidth]{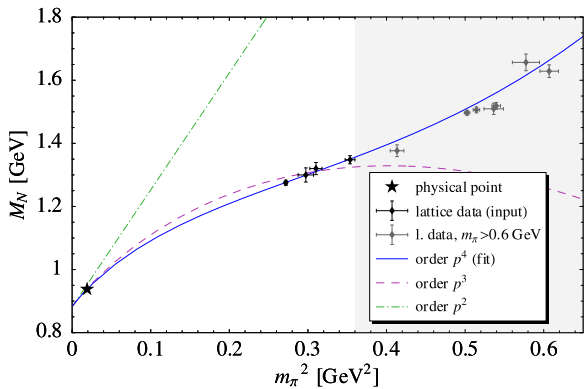}
	\caption{Nucleon mass as function of $m_\pi^2$. Shown is the best fit interpolation between lattice results and the physical point (star), performed at chiral order $p^4$ using eq.\,(\ref{massp4}) (solid curve), with input given in column (a) of table~\ref{tab-paramsb}. The physical point is included. The lattice data points in the grey region have not been used as input. Also shown are intermediate steps at orders $p^2$ and $p^3$ according to eqs.\,(\ref{massp2}),\,(\ref{massp3}). The parameter $e_1^{(3)}(\lambda)$ has been fitted to lattice data.
	}
	\label{fig-conv}
\end{figure}

Using $c_1$ and $M_0$ as given by the $M_N^{(4)}$ best fit in table \ref{tab-paramsb}, we display, in fig.\,\ref{fig-conv}, $M_N^{(2)}$ and examine $M_N^{(3)}$ of eq.\,(\ref{massp3}). The pattern of successive steps is certainly promising for $m_\pi < 600 \units{MeV}$. 
Note that the parameter $e_1^{(3)}(\lambda)$ in $M_N^{(3)}$ and $e_1^{(4)}(\lambda)$ in $M_N^{(4)}$ have different $\lambda$-dependence. 
In contrast to ref.\,\cite{PHW}, we do not identify the two parameters at a specific scale. Instead $e_1^{(3)}(\lambda)$ is fitted to lattice data, yielding $e_1^{(3)}(1\units{GeV}) \simeq 0.48\units{GeV^{-3}}$, a naturally sized value for that coupling. 
A more stringent test of convergence would have to involve an estimate of corrections from ${\cal O}(p^5)$. Then, however, the number of poorly constrained parameters would become prohibitively large. In the present work a consistency check of the low-energy constants required by the fit to the nucleon mass in comparison with the same constants deduced from the analysis of pion-nucleon scattering data provides a non-trivial test that  will be performed in section \ref{sec-empconstr}. \par

\begin{figure}[tb]
	\centering
	\includegraphics[width=\columnwidth]{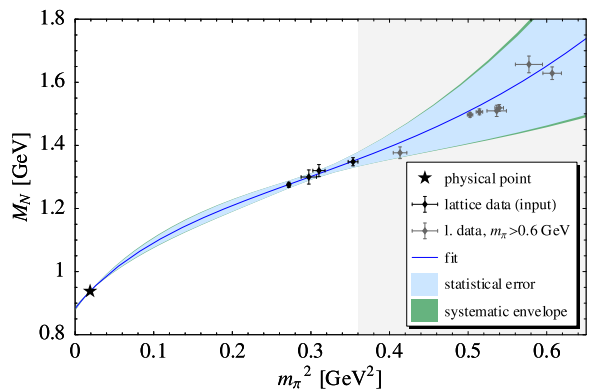}
	\caption{Error band at 68\% confidence level (``statistical error'') and envelope of bands encoding input parameter uncertainties (``systematic envelope'').} 
	\label{fig-band}
\end{figure}

Let us now study the numerical impact of input uncertainties for eq.\,(\ref{massp4}). The relevant technical details are summarized in appendix~\ref{sec-detailserror}.
The uncertainty in $c_3$ is substantial but difficult to quantify. 
For the moment, we ignore uncertainties in $c_2$ and $c_3$ and defer this part of the discussion to sections \ref{sec-empconstr} and \ref{sec-voldep}. To start with, $g_A$ and $f_\pi$ are set equal to their values
at the physical point. Fitting with this setup has 
led to the results of ``fit II'' in \cite{PHW}. We now also take into account uncertainties in extracting the lattice pion mass, see appendix \ref{sec-errpionmass}. They turn out to have little effect. The resulting parameter values are listed in column (a) of table~\ref{tab-paramsb}.
The global error band for the pion mass dependence of the nucleon mass is the ``statistical band'' displayed in fig.\,\ref{fig-band}.

The values of $g_A$ and $f_\pi$, to be taken in the chiral limit,
are expected to differ slightly from the values at the physical point.
We assume to find these values in the intervals
\begin{equation}
	g_A=  1.1\ ...\ 1.3 , \qquad
	f_\pi=  86.2\units{MeV}\ ...\ 92.4\units{MeV}~~.
	\label{eq-gafpiinterv}
	\end{equation}
The range for $g_A$ is taken from \cite{HPW}. For the lower boundary of $f_\pi$, we
take the estimate from an analysis of the pion mass dependence of $f_\pi$ in ref.\,\cite{CD}.
Varying $g_A$ and $f_\pi$ within these intervals extends the error band
to what is labeled ``systematic envelope'' in fig.\,\ref{fig-band} (see appendix \ref{sec-detailserror}).
The corresponding parameter bounds can be found in column (b) of table~\ref{tab-paramsb}. \par

\begin{table*}[tb]
	\caption{Input and output parameters in the fits to large volume lattice data. Lattice pion mass errors have been taken into account. The pion-nucleon sigma term is obtained using $\sigma_N = m_\pi^2\ (\partial M_N / \partial m_\pi^2)$ with eq.\,(\ref{massp4}).
	}
	\label{tab-paramsb}
\begin{ruledtabular}
\begin{tabular}{ll||cc|cc} 
	 & & \multicolumn{2}{l|}{(a) statistical error} & \multicolumn{2}{l}{(b) systematic envelope}  \\
	\hline
	$e_1^{(4)}(1 \units{GeV})$ & $(\mathrm{GeV^{-3}})$  &  0.74 $\pm$ 0.18 & fitted & 0.27 $\dots$ 1.14 & fitted \\   
	$M_0$ & $(\mathrm{GeV})$                            &  0.882 $\pm$ 0.003 & fitted & 0.876 $\dots$ 0.888 & fitted \\
	$c_1$ & $(\mathrm{GeV^{-1}})$                       & $-0.93$ $\pm$ 0.04 & elim. & $-1.04$ $\dots$ $-0.82$ & elim. \\     
	$g_A$ &                                             & 1.267 & fixed & 1.1 $\dots$ 1.3 & scanned \\         
	$f_\pi$ & $(\mathrm{MeV})$                          & 92.4 & fixed  & 86.2 $\dots$ 92.4 & scanned \\    
	$c_2$ & $(\mathrm{GeV^{-1}})$                       & 3.2 & fixed & 3.2 & fixed \\                            
	$c_3$ & $(\mathrm{GeV^{-1}})$                       & $-3.4$ & fixed & $-3.4$ & fixed \\ 
	\hline
	$\chi^2/\text{d.o.f.}$ &                            & 0.13 & & 0.1268 $\dots$ 0.1346 & \\         
	\hline
	$\sigma_N$ & $(\mathrm{MeV})$                       & 49 $\pm$ 3 & & 44 $\dots$ 54 & \\  
	\end{tabular}
	\end{ruledtabular}
\end{table*}

In the limit $m_\pi \rightarrow 0$, the error band remains narrow,
reflecting our ability to determine $M_0$ quite precisely.
Of course, this prediction relies heavily on the inclusion of the physical point
which is located close to the chiral limit.
For pion masses below $0.6\units{GeV}$, i.e. left of the four selected lattice points,
the band does not bulge much. As long as information about the physical point
and the LECs is included, $\chi$PT provides a stable interpolant which is well-conditioned to make predictions within its
range of validity.\footnote{We refer to section 7 for a scenario without inclusion of the physical point.} 
However, the predictive power of our analysis 
in the region $m_\pi > 600\,{\rm MeV}$ is evidently low.
The agreement between best-fit curve and lattice data up to $m_\pi \approx 750\units{MeV}$, 
a scale not small compared to the chiral symmetry breaking scale $\Lambda_\chi \approx 4\pi f_\pi$, 
has low statistical significance.\footnote{Complementary results for large quark masses are obtained using different methods, such as the Adelaide approach \cite{LTY} and the chiral quark soliton model \cite{CQSM}.}

Note that while the individual fit parameters in table~\ref{tab-paramsb} exhibit appreciable sensitivity to the input choice for $g_A$ and $f_\pi$, the overall {\em shape} of the fit curve shows hardly any dependence on $g_A, f_\pi$ in the range of interest for $m_\pi$, such that the ``systematic envelope'' in fig.\,\ref{fig-band} for the global error band is barely distinguishable from the band found for fixed $g_A$ and $f_\pi$. 

In ref.\,\cite{src-Beane} an attempt was made to determine the impact of higher-order effects
by plotting a band for $M_N^{(4)}$,
selecting the chiral limit values of $g_A$ and $f_\pi$ from error intervals
while all other parameters were kept fixed. However,
when studying higher-order effects one must substitute 
the {\em pion mass dependent} functions $g_A(m_\pi)$ and $f_\pi(m_\pi)$, constrained in such a way
that they are consistent with lattice QCD results over the whole range of $m_\pi$. 
This has not been done and therefore the significance of the analysis \cite{src-Beane} remains doubtful. 

Our conclusion is instead that the interpolation based on ${\cal O}(p^4)$ chiral perturbation theory is sufficiently stable for $m_\pi \le 0.6$ GeV.

\section{Consistency with pion-nucleon scattering} 
\label{sec-empconstr}

In order to test whether our extrapolations from lattice results are physically meaningful, we compare them with pion-nucleon observables for which chiral expansions involve the same parameters but in different combinations.
One such observable is the isospin-even $\pi N$ S-wave scattering amplitude at threshold,
$T^+(m_\pi) \equiv  4\pi \left( 1 + m_\pi/M_N \right) a^+ $ where $a^+$ is the corresponding scattering length. Empirically \cite{ELT} one finds the anomalously small value $T^+ = (-0.12 \pm 0.11) \units{GeV^{-1}}$.
Chiral symmetry implies that the leading term of $T^+$ at chiral order $p$ vanishes. In 
HB$\chi$PT the contributions at order $p^2$ and $p^3$ read \cite{BKMpiN}:
\begin{align}
	T^+
	& = \frac{2 m_\pi^2}{f_\pi^2}\left(c_2+c_3-2c_1-\frac{g_A^2}{8 M_0}\right) 
	+ \frac{3 g_A^2 m_\pi^3}{64 \pi f_\pi^4} \nonumber\\
	& + {\cal O}(p^4)~~.
	\label{eq-tplus}
	\end{align}
The term of order $p^4$, supposedly small, involves a series of additional low-energy constants which are not well determined and therefore presently of no practical use for a detailed estimate. 

Another suitable observable is the isospin-even, spin-averaged P-wave $\pi N$ threshold amplitude, 
$P_1^+(m_\pi) \equiv  4\pi ( 1 + m_\pi/M_N ) (4 a_{33} + 2 a_{31} + 2 a_{13} + a_{11} )$, given in terms
of the P-wave scattering volumes $a_{2I,2J}$ in channels with spin/isospin $J, I =$ 1/2 or 3/2. 
Empirically, one finds $P_1^+ = (1044 \pm 38) \units{GeV^{-3}}$ in \cite{src-p1plusexp}.
In HB$\chi$PT, the terms up to chiral order $p^3$ are \cite{BKMpiN}:
\begin{align}
	P_1^+
	& = \frac{2}{f_\pi^2} \left( c_2 \frac{m_\pi}{M_0} - c_3 \right) + \frac{g_A^2 m_\pi}{4 f_\pi^2 M_0^2} - \frac{g_A^2 m_\pi}{12 \pi f_\pi^4} \left( g_A^2 + \frac{77}{32} \right) \nonumber\\
	& + {\cal O}(p^4)~~,
	\label{eq-poneplus}
\end{align}
where one should note that $P_1^+$ enters in the scattering amplitude with an extra factor involving pion momenta, $\vec{q}\cdot 
\vec{q}\,'$, of order $p^2$ in the chiral counting.

Further independent information can be drawn from the pion-nucleon sigma term,
\begin{equation}
\sigma_N = \langle N|m_u\,\bar{u}u + m_d\,\bar{d}d|N\rangle = m_q{\frac{\partial M_N}{\partial m_q}}~~.
\end{equation}
Making use of the Gell-Mann--Oakes--Renner relation, one obtains:
\begin{equation}
\sigma_N \simeq m_\pi^2{\frac{\partial M_N}{\partial m_\pi^2}} = - 4c_1m_\pi^2 + \dots
\label{sigma}
\end{equation}  
Ref.\,\cite{GLS} provides us with the empirical value $\sigma_N = (45 \pm 8)$ MeV. 

The requirement that the empirical $T^+$, $P_1^+$and $\sigma_N$ are simultaneously reproduced implies $c_3 \simeq -(5 \pm 1)\units{GeV^{-1}}$ if one uses the $\mathcal{O}(p^3)$ HB$\chi$PT formulae eqs.\,(\ref{eq-tplus}),\,(\ref{eq-poneplus}) at the physical pion mass.
It might then appear that $c_3$ determined from $\pi N$ scattering is incompatible with the value $c_3 = -3.4 \units{GeV^{-1}}$ consistent with $NN$ scattering and used as input previously. 
At tree level and in the non-relativistic limit, the spin-isospin averaged P-wave scattering volume is well known \cite{BW,EW} to be proportional to $(\Delta^2 - \omega^2)^{-1}$, where the pion energy $\omega$ equals $m_\pi$ at threshold. The fact that the delta-nucleon mass difference, $\Delta = M_\Delta - M_N$, is barely twice the physical pion mass, $m_\pi^\text{phys}$, is at the origin of the strong energy dependence of the P-wave $\pi N$ amplitude. 
A determination of $c_3$, first by comparison with $\pi N$ threshold data and secondly by examining its role in peripheral $NN$ phase shifts, will therefore lead to different conclusions concerning $c_3$, the $NN$ situation being effectively closer to the limit of a static pion field with $\omega = 0$ \cite{KGW}. This apparent discrepancy just reflects the in-effectiveness of HB$\chi$PT to deal with the well-known strong energy dependence of the $\pi N$ amplitude. The important effects of this energy dependence are then ``hidden'' by absorbing a large correction of order $m_\pi^2/\Delta^2$ into $c_3$ when using the $\mathcal{O}(p^3)$ result of eqs.\,(\ref{eq-tplus}),\,(\ref{eq-poneplus}) for its determination. Including this leading correction the estimate of $c_3$ is reduced by a factor of $1-(m_\pi/\Delta)^2 \approx 3/4$, which accounts for much of the difference between the two cases. The large value of $c_3\approx -5 \units{GeV^{-1}}$ is therefore an artifact of the HB$\chi$PT expansion truncated at ${\cal O}(p^3)$.

Matching the tree-level $\pi N$ amplitude with explicit $\Delta(1232)$ and the $\pi N \to \pi N$ graph calculated from the second-order pion-nucleon effective Lagrangian, the $\Delta(1232)$ contribution to $c_3$ reads
\begin{equation}
c_3^\Delta = - \frac{4\,c_A^2}{9\,\Delta}\,\,,
\end{equation}
where $c_A = g_{\pi N \Delta}\,f_\pi/(2M_0)$ in terms of the $\pi N \Delta$ coupling constant $g_{\pi N \Delta}$, and all constants $(f_\pi, \Delta, M_0)$ are understood to be taken in the chiral limit. With the frequently used empirical coupling $c_A\simeq 1.5$, see section \ref{sec-deltanumres}, we have $c_3^\Delta \simeq -3.4 \units{GeV^{-1}}$, which agrees with $c_3$ extracted from the $NN$ scattering analysis. We interpret this value as being representative for the $c_3$ to be used in the extrapolation of lattice results for $M_N$. 

We point out here that the set of values $M_0=0.883\units{GeV}$, $c_1=-0.89\units{GeV^{-1}}$, $c_2=2.98\units{GeV^{-1}}$, $c_3=-3.55\units{GeV^{-1}}$, $e_1^{(4)}(1\units{GeV}) = 0.46\units{GeV^{-3}}$, $g_A = 1.1$ and $f_\pi=92.4\units{MeV}$ represents an optimal fit to lattice data, gives a curve right in the center of our statistical error band fig.\,\ref{fig-band}, and satisfies the empirical constraints for $\sigma_N$, $T^+$ and $P_1^+$ {\it provided that $c_3$ in eqs.\,(\ref{eq-tplus}),\,(\ref{eq-poneplus}) is replaced by $(4/3) c_3 \equiv c_3^{\pi N}$}.
We note that the factor $4/3$ is to be understood as representing the substantial contribution at ${\cal O}(p^4)$ in eqs.\,(\ref{eq-tplus}),\,(\ref{eq-poneplus}) arising from the $\Delta(1232)$ propagator structure.
Our intermediate conclusion is that the parameters required by the ``best fit'' to lattice results for $M_N$ are consistent with those from the analysis of $\pi N$ and $NN$ low-energy data once $c_3$ is corrected for artifacts of the $p^3$ Heavy Baryon truncation.  
These considerations will be strengthened as we now further examine the role of the $\Delta(1232)$ and investigate the systematics of finite volume effects.

\section{Explicit $\Delta\,(1232)$ degrees of freedom} 
\label{deltasec}

An effective field theory which includes only pion and nucleon as explicit degrees of freedom encodes implicitly contributions from the $\Delta(1232)$ resonance through low-energy constants, e.g. $c_3$, but its interpretation requires caution, as elaborated in the previous section. Working at limited perturbative order, ``freezing'' the $\Delta(1232)$ and relegating its effects to higher order terms can lead to a rather in-effective chiral expansion, given that the delta-nucleon mass difference is only about a quarter of the chiral symmetry breaking scale, $\Lambda_\chi \approx 4 \pi f_\pi$. It has been shown that including the $\Delta(1232)$ as an {\it{explicit}} degree of freedom in spin-dependent quantities like the magnetic moments or the axial coupling of the nucleon promotes important quark-mass dependent contributions to low orders in the perturbative calculation, leading to well-behaved chiral extrapolation functions, see e.g. refs.\,\cite{HW,HPW}. For the case of the nucleon mass explicit $\Delta$(1232) treatment turns out to be less crucial. However, it helps us in clarifying the role of the $p^4$-effects in eq.\,(\ref{massp4}), which are dominated by the couplings $c_i$.

\subsection{Formalism}

We work in the Lorentz covariant formulation of the so-called small scale expansion (SSE) introduced in ref.\,\cite{BHMdelta}. In this scheme the delta-nucleon mass difference in the $SU(2)$ chiral limit, $\Delta = M_\Delta - M_0$, is included in the power-counting, together with the pion mass and soft external momenta, as a small scale generically labeled $\epsilon$ \cite{HHK}. 

\begin{figure}[t]
  \begin{center}
    \includegraphics*[width=0.25\columnwidth]{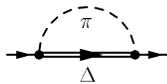}
    \caption{Leading-one-loop diagram for the nucleon self-energy with an intermediate $\Delta\,(1232)$.}
   \label{massdeltagraph}
  \end{center}
\end{figure}

The main topic of our discussion with explicit delta degrees of freedom is to clarify the success of the ${\cal O}(p^4)$ calculation without $\Delta(1232)$. The leading effect due to explicit delta appears at order $\epsilon^3$ in SSE and starts to contribute at $m_\pi^4$, like the ${\cal O}(p^4)$ graphs.
Therefore we have concentrated on the ${\cal O}(\epsilon^3)$ leading one-loop contribution to the nucleon self-energy involving the propagation of the $\Delta(1232)$, fig.\,\ref{massdeltagraph}.\footnote{For a leading one-loop calculation in a different framework see ref.\,\cite{Pascal}.} In Section \ref{eps4} we will compare our $\epsilon^3$ results with the $\epsilon^4$ expression in ref.\,\cite{BHMdelta2} in order to estimate the importance of higher-order effects.

The $\epsilon^3$ calculation requires the leading chiral pion-nucleon-delta effective Lagrangian
\begin{equation}
{\cal{L}}_{\pi N \Delta}\,=\,-\frac{c_A}{f_\pi}\,\bar{\Psi}^i_\mu \partial^\mu \pi^i \Psi_N +{\rm{h.c.}} ~~,
\end{equation}
where ${\Psi}^i_\mu$ is the Rarita-Schwinger field representing the $\Delta(1232)$ and $\Psi_N$ is the nucleon field.
We employ the following propagator for the free spin-3/2 isospin-3/2 field \cite{BHMdelta2}:
\begin{eqnarray}
\mathcal{G}_{\mu\nu}^{i j} (p)= -i\,\frac{p\!\!\!/ + M_\Delta}{ p^2 - M_\Delta^2+i\epsilon}\,\frac{p^2}{M_\Delta^2}\, P_{\mu\nu}^{3/2}\,\xi^{i j}_{3/2}~, \label{prop2}
\end{eqnarray} 
where $P_{\mu \nu}^{3/2}$ and $\xi^{i j}_{3/2}$ are the spin- and isospin-3/2 projection operators, respectively.
Using the propagator (\ref{prop2}), the computation of the one-loop $\Delta(1232)$ contribution to the nucleon self-energy in infrared regularization requires the standard loop integrals {\footnote{See ref.\,\cite{BL} for details. In distinction from this reference, we do not identify the regularization scale with $M_0$, which is a parameter in our approach.}}
\begin{align}
iI_\Delta (p^2)&= \int \frac{d^d k} {(2\pi)^d} \frac{1}{ (m_\pi^2 - k^2-i \epsilon)
[M_\Delta^2 - (p-k)^2-i \epsilon]}~, \nonumber\\
i\Delta_\pi&=\int \frac{d^d k} {(2\pi)^d} \frac{1}{(m_\pi^2- k^2 -i \epsilon)}~. \nonumber
\end{align}
Here $d$ denotes the space-time dimension. Both previous integrals diverge as $d \to 4$. 

The ${\cal O}(\epsilon^3)$ graph in fig.\,\ref{massdeltagraph} leads to a correction to the mass of the nucleon, cf. \cite{BHMdelta2}: 
\begin{eqnarray}
\delta M_N^{\Delta}& = &\frac{- c_A^2\,(d-2)}{4 {f_\pi}^2 (d-1)\,  M_0\, M_\Delta^2} 
\Bigg\{I_\Delta(M_0^2)\,(M_0^2-2 M_\Delta M_0 \nonumber \\ & & +\,  M_\Delta^2-m_\pi^2)(M_0^2+2M_\Delta M_0+M_\Delta^2-m_\pi^2)^2 \nonumber\\ & & +\, \Delta_\pi\Big[M_0^2\left(M_0^2+2M_\Delta M_0 +4 m_\pi^2\,\frac{d-1}{d}\right) \nonumber \\ & & -\, (M_\Delta^2 - m_\pi^2)(2M_\Delta M_0 +M_\Delta^2-m_\pi^2)\Big]\Bigg\} 
\label{mshifteps3}
\end{eqnarray}
where $d=4$ is set after removal of the singularities from $I_\Delta$ and $\Delta_\pi$.
The resulting expression for the nucleon mass at order $\epsilon^3$ is then of the form
\begin{align}
M_N&=M_0-4\,c_1 m_\pi^2+\delta M_N^{\pi N}+\delta M_N^{\Delta}\nonumber\\&+{\rm{counter\, terms}}~, \label{massaeps3} 
\end{align}
where $\delta M_N^{\pi N}$ is the nucleon mass-shift at order $p^3$ in the scheme with $\pi N$ degrees of freedom only, cf. eq.\,(\ref{massp3}).
We note that we recover the non-relativistic ${\cal O}(\epsilon^3)$ SSE result for the nucleon mass\footnote{See, e.g., the calculation in ref.\,\cite{FF}, which in addition contains an ``effective'' coupling, like $e_1^{(3)}$ in section~\ref{sec-nucmass}.} by just keeping the leading term in the $1/M_0$ expansion in eq.\,(\ref{massaeps3}) with eq.\,(\ref{mshifteps3}) inserted.

In a first approach we truncate at order $1/M_0$ the result of eq.\,(\ref{massaeps3}). In addition to the term $-4\,e_\Delta\,m_\pi^4\,\bar \Psi_N \Psi_N$, the SSE-scheme provides two more counter terms at this order\footnote{The constants $B_1$ and $B_2$ used here are denoted $B_{32}$ and $B_{23}$, respectively, in ref.\,\cite{BHMdelta2}.}
\begin{equation}
{\cal L}_{\pi N}^{(3)}\,=\,\bar{\Psi}_N\left[B_1\,\Delta^3+ 4B_2\,m_\pi^2\,\Delta\right]\Psi_N + \dots \label{Bi} 
\end{equation}
which are sufficient for renormalization. The finite parts of the renormalized couplings $B_{1,2}$ are then set such that the chiral expansion of $M_N(m_\pi)$ at order $\epsilon^3$ begins with
\begin{equation}
M_N=M_0-4\,c_1 m_\pi^2+{\cal O}(m_\pi^3)~. \label{decconstr}
\end{equation} 
Imposing this condition, one obtains
\begin{align}
B_{1}^r(\lambda)&=\frac{8\,c_A^2}{9 (4 \pi {f_\pi})^2}\left[2 \left(1-\frac{\Delta}{M_0}\right)-3 \left(2+\frac{\Delta}{M_0}\right)\ln{\frac{2\Delta}{\lambda}}\right]\nonumber\\
B_{2}^r(\lambda)&=\frac{2\,c_A^2}{(4 \pi {f_\pi})^2}\left[\frac{\Delta}{M_0} +\left(2+\frac{\Delta}{M_0}\right)\ln{\frac{2\Delta}{\lambda}}\right]~.
\end{align}
It follows that
\begin{align}
M_N=\;&M_0-4\,c_1 m_\pi^2-\frac{3g_A^2 m_\pi^3}{32 \pi f_\pi^2 }-\frac{3\,m_\pi^4}{64\,\pi^2 f_\pi^2\, M_0}\,(g_A^2+3c_A^2)\nonumber\\ &-\frac{m_\pi^4}{32 \pi^2 f_\pi^2\, M_0}\,(3g_A^2+10 c_A^2) \ln{\frac{m_\pi}{\lambda}}+4\,e_\Delta(\lambda) m_\pi^4\nonumber\\ &-\frac{c_A^2\,m_\pi^2 \Delta}{12 \pi^2 f_\pi^2}\left(1+\frac{\Delta}{2 M_0}\right)\nonumber\\&-\frac{c_A^2}{6 \pi^2 f_\pi^2}\, (2\Delta^3 - 3 m_\pi^2 \Delta)\left(1+\frac{\Delta}{2 M_0}\right)\ln{\frac{m_\pi}{2\Delta}}\nonumber\\ &-\frac{c_A^2}{3 \pi^2 f_\pi^2 } \left(1+\frac{\Delta}{2 M_0}\right)(\Delta^2-m_\pi^2)^{3/2}\nonumber\\ &\times \ln{\left(\frac{\Delta}{m_\pi}+\sqrt{\frac{\Delta^2}{m_\pi^2}-1}\,\right)}
\label{massSSEtrunc} 
\end{align} 
The last formula is valid for $m_\pi \leq \Delta$. Since in our numerical analysis we take as input lattice data at pion masses larger than the physical one, we need also the analytic continuation of the expressions above to $m_\pi \geq \Delta$. This is done with the replacement
\begin{align} 
&\sqrt{\Delta^2-m_\pi^2}\,\ln\left(\frac{\Delta}{m_\pi}+\sqrt{\frac{\Delta^2}{m_\pi^2}-1}\,\right)\quad\to\quad\nonumber\\ &\;-\sqrt{m_\pi^2-\Delta^2}\arccos\frac{\Delta}{m_\pi}~. \nonumber
\end{align}

In the next section we will also analyse the non-truncated ${\cal O}(\epsilon^3)$ result of eq.\,(\ref{massaeps3}) with eq.\,(\ref{mshifteps3}), keeping the full tower of $1/M_0^n$ recoil corrections. We introduce the same counter term structures as for the truncated expression and accept uncompensated regularization scale dependence $\sim m_\pi^6$ and $\sim m_\pi^8$ in this case. The finite parts of the renormalized $B_i$ are again determined by imposing eq.\,(\ref{decconstr}).

\subsection{Numerical results}
\label{sec-deltanumres}

Using as input our selected set of lattice data discussed in section \ref{sec-nucmass}, we now analyze the ${\cal{O}}(\epsilon^3)$ expressions for $M_N(m_\pi)$ introduced in the previous section: the ${\cal O}(1/M_0)$-truncated eq.\,(\ref{massSSEtrunc}) and the ``full'' expression with its residual scale dependence.
We fix the renormalization scale again at $\lambda=1\,{\rm{GeV}}$. The mass splitting $\Delta$ is identified with its value at the physical pion mass.
The axial vector nucleon-delta coupling $c_A$ is treated as input quantity, determined from the decay width of the $\Delta(1232)$ as follows.
In the rest frame of the decaying delta we have the relativistic expression:
\begin{equation}
\Gamma_{\Delta \to N \pi}=\frac{c_A^2\,q^3}{6 \pi f_\pi^2}\:{\frac{({M_\Delta}+M_N)^2-m_\pi^2}{4M_\Delta^2}}~,  \label{carel} 
\end{equation}
where $q$ is the pion or nucleon momentum in that frame. At the delta mass and width corresponding to its pole position  \cite{PDG04}, $M_\Delta - i\,\Gamma_\Delta/2 = (1210 - i\,50)$ MeV, one finds $c_A \simeq 1.5$. 
We fit again the three remaining unknown parameters: $M_0,\,c_1$ and $e_\Delta(1$ GeV$)$.

\begin{table*}[tb]
 \caption{Fit results for $M_N(m_\pi)$ in covariant SSE, at leading-one-loop order. ``Fit delta I'' refers to the formula truncated at order $1/M_0$, ``fit delta II'' to the ``full'' expression with residual scale dependence. The input parameters are $c_A = 1.5, g_A = 1.267, f_\pi = 92.4$ MeV and $\Delta = 271.1$ MeV. The renormalized effective coupling $e_\Delta$ entering at ${\cal O}(m_\pi^4)$ has a different regularization scale dependence in the two cases.
 \label{tabledelta}}
  \begin{ruledtabular}
  \begin{tabular}{c||cccc} 
      & $M_0\;({\rm GeV})$ & $c_1\;({\rm GeV}^{-1})$ & $e_\Delta(1\,{\rm{GeV}})\;({\rm GeV}^{-3})$ & ${\chi^2}/{\rm{d.o.f.}}$    \\ \hline 
 fit delta I &  $0.894 \pm 0.004$ & $-0.76 \pm 0.05$ & $4.5 \pm 0.1$ & 0.19 \\ 
 fit delta II &  $0.873 \pm 0.004$ & $-1.08 \pm 0.05$ & $2.8 \pm 0.2$ & 0.43 \\ 
  \end{tabular}
  \end{ruledtabular}
\end{table*}

\begin{figure}[tb]
  \begin{center}
    \includegraphics*[width=\columnwidth]{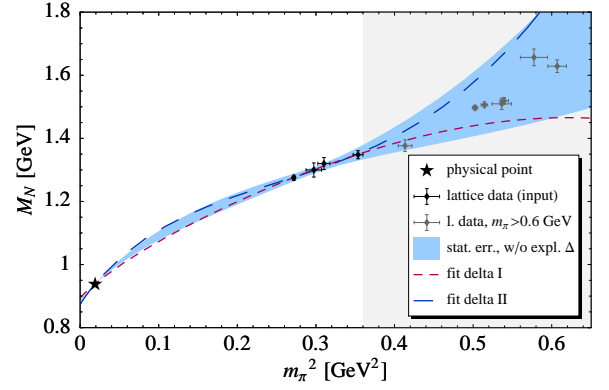}
    \caption{Best fit curves based on the formula at order $\epsilon^3$ in SSE. The short-dashed curve refers to ``fit delta I'', while the long-dashed curve corresponds to ``fit delta II'' in table~\ref{tabledelta}. For comparison, we plot the 68\% statistical error band of the ${\cal O}(p^4)$ B$\chi$PT result shown in fig.\,\ref{fig-band}.}
    \label{massSSEfullfig}
  \end{center}
\end{figure}

In table~\ref{tabledelta}, ``fit delta I'' refers to eq.\,(\ref{massSSEtrunc}), while ``fit delta II'' refers to the relativistic untruncated expression. We fit to the four data with $m_\pi < 600\,{\rm MeV}$ including the physical point, see fig.\,\ref{massSSEfullfig}. 
In ``fit delta II'' the output parameters $M_0$ and $c_1$ remain stable (within error bars) if we vary the input scale $\lambda$ between $0.5$ and $1.5\,{\rm{GeV}}$. Thus the residual scale dependence is indeed under control.
In light of the discrepancies between the two fits, the truncation of eq.\,(\ref{massSSEtrunc}) at order $1/M_0$ neglects terms which seem to play a significant role. Furthermore, we have checked that the $68\%$ error bands associated with ``fit delta I'' and ``fit delta II'' do not overlap for $m_\pi < 300$ MeV, except at the physical point used as a constraint. Therefore we can conclude that the discrepancy between the truncated and untruncated expressions does not have a statistical origin. This situation is different in the ${\cal O}(p^4)$ B$\chi$PT calculation discussed in ref.\,\cite{PHW}, where truncating at $1/M_0$ represents a very good approximation to the full result.

In fig.\,\ref{massSSEfullfig} we plot the curves corresponding to ``fit delta I'' and ``fit delta II'', together with the 68\% error band of the ${\cal O}(p^4)$ B$\chi$PT result in section \ref{sec-erranalysis} previously drawn in fig.\,\ref{fig-band}.  At that confidence level, B$\chi$PT at ${\cal O}(p^4)$ and ${\cal O}(\epsilon^3)$ covariant SSE are compatible for the whole range of pion masses under study. At the present level of accuracy, treating the $\Delta\,(1232)$ as an explicit, propagating field is therefore not essential for a satisfactory description of the quark mass dependence of the nucleon mass. The virtual $\Delta(1232)$, being far off-shell, has a short propagation length. Therefore an equally successful interpolating function can be obtained by ``freezing" the delta effects into low-energy constants, working at fourth chiral order, with a value for the coupling $c_3$ governed by delta-dominance, as discussed in section \ref{sec-empconstr}. Note that the $\pi \Delta$ loop integral in fig.\,\ref{massdeltagraph} involves a denominator $(\sqrt{\vec q^{\ 2} + m_\pi^2} + \Delta)^{-1}$, characteristic of the crossed (u-channel) delta pole in the pion-nucleon scattering amplitude, for which the sensitivity to $\Delta$ turns out to be less significant.


These conclusions hold for any choice of the input parameters $g_A$, $f_\pi$, $c_A$ and $\Delta$ within phenomenologically meaningful limits. In order to check the sensitivity with respect to variations of the delta-nucleon mass difference we have performed fits for $\Delta=293\,{\rm MeV}$ and $330\,{\rm MeV}$. The former corresponds to the $90^0$ $\pi N$ phase-shift in the spin-3/2 isospin-3/2 channel, while the latter is motivated as a chiral limit value in ref.\,\cite{BHMdelta2}.

\subsection{${\cal O}(\epsilon^3)$ vs. ${\cal O}(\epsilon^4)$}
\label{eps4}

\begin{figure}[ht]
  \begin{center}
    \includegraphics*[width=\columnwidth]{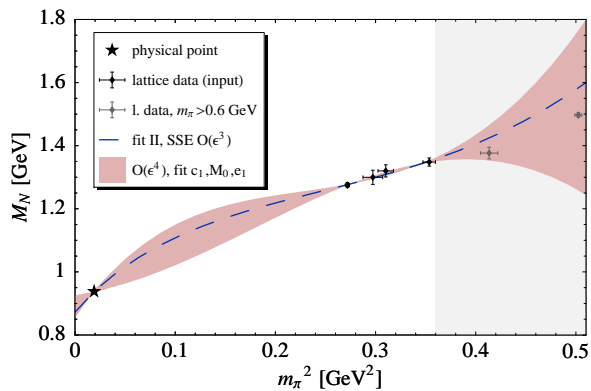}
    \caption{Comparison between our ${\cal O}(\epsilon^3)$ ``fit delta II'' (dashed curve) and a band of ${\cal O}(\epsilon^4)$ fits using natural size assumptions for higher-order couplings.}
    \label{mneps4}
  \end{center}
\end{figure}

Eq.\,(20) of ref.\,\cite{BHMdelta2} provides the $\epsilon^4$ contribution to $M_N(m_\pi)$.\footnote{The reader should be aware of misprints in eqs.\,(19), (25) and (26) of that paper.} This expression involves the poorly known SSE couplings $c_2$, $c_3$, $b_3$, $b_6$ and $e_1$. Note that $c_2$, $c_3$ and $e_1$ have different numerical values in SSE and B$\chi$PT. 
Motivated by delta-dominance, we assume the range $-1 \dots 1\,{\rm GeV}^{-1}$ for $c_2$ and $c_3$ within SSE. For $b_3$ and $b_6$ we have made a more conservative estimate, varying them between $-3 \dots 3\,{\rm GeV}^{-1}$. Scanning those ranges, we have performed fits with three free parameters $M_0$, $c_1$ and $e_1(\lambda=1\,{\rm GeV})$. We have fixed $g_A=1.267$, $f_\pi=92.4\,{\rm MeV}$, $\Delta=271.1\,{\rm MeV}$ and $c_A=1.5$. The resulting band of best-fit curves is shown in fig.\,\ref{mneps4}. In output we get: $M_0=0.85 \dots 0.93\,{\rm GeV}$, $c_1=-1.32 \dots{-0.25}\,{\rm GeV^{-1}}$ and $e_1(1\,{\rm GeV})=-7.3 \dots 1.9\,{\rm GeV}^{-3}$. In view of these results, our $\epsilon^3$ results are potentially affected by important systematic effects from higher orders. At present, however, lack of information on SSE couplings does not permit us to perform the analysis discussed in previous chapters at ${\cal O}(\epsilon^4)$.


\section{Volume Dependence}
\label{sec-voldep}

\subsection{Implementing finite lattice sizes}

So far we had to restrict ourselves to the largest available lattice sizes. However, $\chi$PT is also able to describe deviations from the infinite-volume limit due to the finite spatial extent of the simulation volume  $L^3$ \cite{GLfv}. Combining the $L$- and $m_\pi$-dependence in our analysis of the nucleon mass enables us to enlarge the input data base for our analysis.

\begin{figure}[ht]
  \begin{center}
    \includegraphics*[width=\columnwidth]{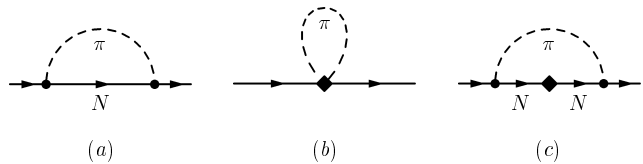}
    \caption{One-loop graphs of ${\cal O}(p^3)$ (a) and ${\cal O}(p^4)$ (b, c) contributing to the nucleon self-energy in B$\chi$PT (without explicit $\Delta$(1232) degrees of freedom). The solid dot (diamond) denotes a vertex from first (second) order $\pi N$ Lagrangian.}
    \label{diags}
  \end{center}
\end{figure}

With pion loop momenta restricted to the discrete values permitted by periodic
boundary conditions, the $\mathcal{O}(p^4)$ B$\chi$PT result for the mass difference $\delta M_N$ between finite and infinite volume has been published in ref.\,\cite{QCDSF}. At order $p^3$ one obtains from diagram (a) of fig.\,\ref{diags} using infrared regularization
\begin{align}
	\delta^{(3)}M_N = &\
		\frac{3 g_A^2 M_0 m_\pi^2}{16 \pi^2 f_\pi^2} 
		\left[\int_0^1 dx+ \int_1^\infty dx \right] 
		\sum_{\vec{n}\in\mathbb{Z}^3\backslash\{\vec{0}\}} \nonumber\\
		& K_0 \left(L |\vec{n}| \sqrt{M_0^2 x^2 + m_\pi^2 (1-x)}\right)~, 
		\label{eq-deltamvolp3} 
	\end{align}
where the integral from 1 to infinity is taken in accordance to the prescription in ref.\,\cite{BL}.

The $\mathcal{O}(p^4)$ contribution stems solely from the tadpole graph \cite{QCDSF}, diagram (b) of fig.\,\ref{diags}, and reads 
\begin{align}
	\delta^{(4)}M_N & = \frac{3 m_\pi^4 }{4 \pi^2 f_\pi^2} \sum_{\vec{n}\in\mathbb{Z}^3\backslash\{\vec{0}\}} 
		\bigg( (2 c_1- c_3) \frac{K_1(L |\vec{n}| m_\pi)}{L |\vec{n}| m_\pi} \nonumber\\ & + 
		c_2 \frac{K_2(L |\vec{n}| m_\pi)}{(L |\vec{n}| m_\pi)^2} \bigg)~. 
		\label{eq-deltamvolp4}
	\end{align}
Here $K_0$, $K_1$ and $K_2$ are modified Bessel functions. Eqs.\,(\ref{eq-deltamvolp3}),\,(\ref{eq-deltamvolp4}) have been worked out in the so called ``$p$-r\'egime'', namely for $m_\pi L \gg 1$.  
In ref.\,\cite{QCDSF} the parameters of eq.\,(\ref{massp4}) have been fitted for large volumes, $L \simeq 2\units{fm}$. The resulting predictions for the finite volume effects at $\mathcal{O}(p^4)$ showed a surprisingly good agreement with dynamical improved Wilson data of the QCDSF, UKQCD and JLQCD collaborations, even down to  $L\approx 1\units{fm}$. Here we use eqs.\,(\ref{eq-deltamvolp3}),\,(\ref{eq-deltamvolp4}) for a global fit of lattice data in different simulation volumes.

We note that no new parameters enter in the finite-volume corrections $\delta M_N$, and that $c_1$, $c_2$ and $c_3$ appear in combinations
different from those in the infinite volume formula (\ref{massp4}).
Thus the $m_\pi$- and $L$-dependence complement each other in constraining these parameters.
The nucleon mass in the finite volume reads
\begin{align}
	M_N(m_\pi,L) & = M_N(m_\pi) \nonumber\\ & + \delta^{(3)}M_N(m_\pi,L) + \delta^{(4)}M_N(m_\pi,L)~,
	\label{massvol}
	\end{align} 
where $M_N(m_\pi)$ is the order $p^4$ result of eq.\,(\ref{massp4}), corresponding to the limit $L \rightarrow\infty$. For the numerical evaluation of $\delta M_N(m_\pi,L)$, it is precise enough to terminate the infinite sums at $|\vec{n}|=8$ and $|\vec{n}|=6$ in eqs.\,(\ref{eq-deltamvolp3}) and (\ref{eq-deltamvolp4}), respectively.
The integral in eq.\,(\ref{eq-deltamvolp3}) is computed numerically.\par

\begin{figure}[tb]
	\centering
	\includegraphics[width=\columnwidth]{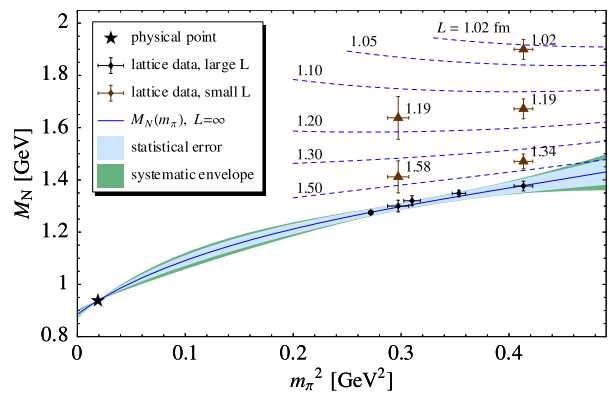}
	\caption{Finite volume lattice data and inferred nucleon mass $M_N(m_\pi,L)$. All lattice points shown here are used as input for the fit, cf. table in appendix \ref{sec-latticedat}. 
	The solid curve, the $68\%$ statistical error band and the systematic envelope project the full information, including the finite volume data, onto the infinite volume limit, eq.\,(\ref{massp4}). The calculated finite volume dependence is shown by the dashed curves. }
	\label{fig-volband}
\end{figure}

\begin{figure}[tb]
	\centering
	\includegraphics[width=0.7\columnwidth]{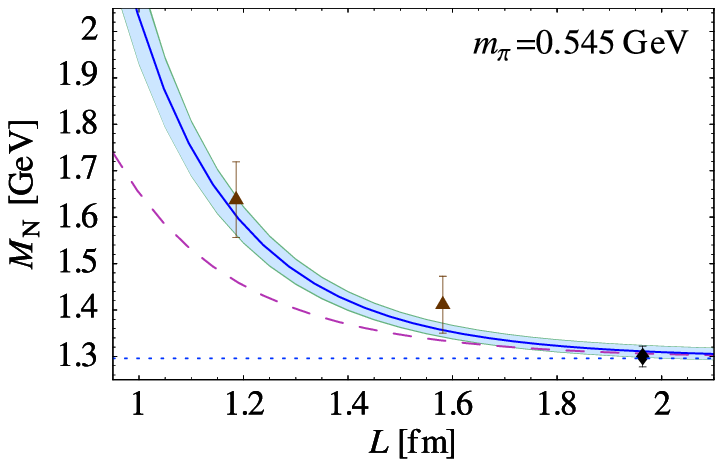} \\ \vspace{5mm}
	\includegraphics[width=0.7\columnwidth]{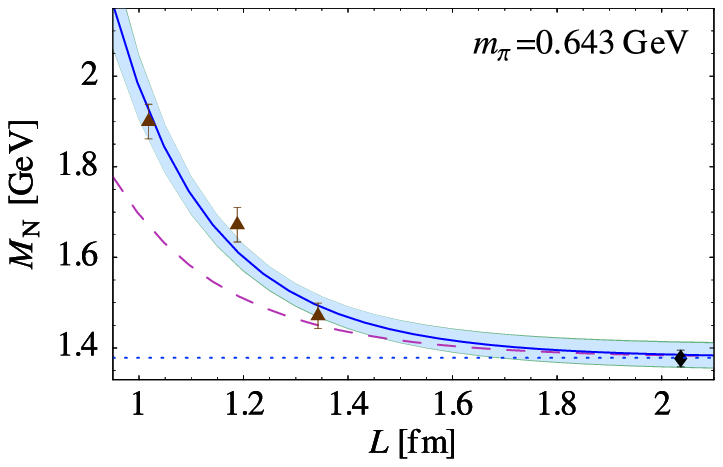}
	\caption{$L$ dependence of the fit function and error bands for fixed pion masses. The horizontal dotted lines are the infinite volume limit of the fit function. The dashed lines represent the $\mathcal{O}(p^3)$ contribution to the finite volume correction.}
	\label{fig-voll}
\end{figure}

Appendix \ref{sec-latticedat} lists the lattice data used for this purpose and explains our selection criteria.
The mass $m_\pi$ in eq.\,(\ref{massvol}) is the pion mass in the infinite volume. For each set of lattice simulations, we have identified the infinite volume pion masses with $m_\pi$ at the largest volume ($L \gtrsim 2\units{fm}$). 
Eq.\,(\ref{massvol}) is used to determine low-energy parameters by fitting to finite volume lattice data following the same statistical strategy as described in section \ref{sec-erranalysis}.
We can now afford to release $c_3$ and determine it from the fit. Also, we accommodate an uncertainty about $c_2$ in a range from $3.1\units{GeV^{-1}}$ to $3.3\units{GeV^{-1}}$, which encompasses results from several HB$\chi$PT fits at $\mathcal{O}(p^3)$
to experimental $\pi N$ scattering data, see table 4 in \cite{FM}. 
The output parameters listed in table \ref{tab-paramsvol} are then used to draw 
\begin{itemize}
	\item	best-fit curve and $68\%$ confidence level error bands based on the infinite-volume expression eq.\,(\ref{massp4}), see fig.\,\ref{fig-volband}.
	\item $L$-dependence at fixed $m_\pi$ based on eq.\,(\ref{massvol}), see fig.\,\ref{fig-voll}.
\end{itemize}
Remarkably, $c_3$ comes out low in magnitude, compatible with $N N$ scattering results \cite{EM}, the outcome of ref.\,\cite{PHW}, the finite volume study of ref.\,\cite{QCDSF} and delta-dominance.
In the previous infinite-volume analysis, it could be argued that $c_3$ and the counter term parameter $e_1^{(4)}(\lambda)$ can be played against each other, given that $e_1^{(4)}$ remains essentially unconstrained. The $L$-dependence now provides a further condition which definitively points to the ``small'' $c_3$ discussed previously in section \ref{sec-empconstr}.

\begin{table*}[tb]
	\caption{Input and output parameters for the finite volume fit.}
	\label{tab-paramsvol}
\begin{ruledtabular}
\begin{tabular}{ll||cc|cc} 
	& & \multicolumn{2}{l|}{(a) statistical error} & \multicolumn{2}{l}{(b) systematic envelope} \\
	\hline
	$e_1^{(4)}(1 \units{GeV})$ & $(\mathrm{GeV^{-3}})$  & 0.60 $\pm$ 0.09 & fitted & 0.22 $\dots$ 0.90 & fitted \\
	$c_3$ & $(\mathrm{GeV^{-1}})$                       & $-2.9$ $\pm$ 0.6 & fitted & $-4.3$ $\dots$ $-1.4$ & fitted \\
	$M_0$ & $(\mathrm{GeV})$                            & 0.884 $\pm$ 0.006 & fitted & 0.873 $\dots$ 0.898 & fitted \\
	$c_1$ & $(\mathrm{GeV^{-1}})$                       & $-0.88$ $\pm$ 0.09 & elim. & $-1.03$ $\dots$ $-0.69$ & elim. \\
	$g_A$ &                                             & 1.267 & fixed & 1.10 $\dots$ 1.30 & scanned \\
	$f_\pi$ & $(\mathrm{MeV})$                          & 92.4 & fixed & 86.2 $\dots$ 92.4 & scanned \\
	$c_2$ & $(\mathrm{GeV^{-1}})$                       & 3.2 & fixed & 3.1 $\dots$ 3.3 & scanned \\
	\hline
	$\chi^2/\text{d.o.f.}$ &                            & 0.75 & & 0.69 $\dots$ 0.82 & \\
	\hline
	$\sigma_N$ & $(\mathrm{MeV})$                       & 47.6 $\pm$ 4.7 & & 36 $\dots$ 57 &  \\
	\end{tabular} 
	\end{ruledtabular}
\end{table*}

\subsection{Finite size effects with explicit $\Delta(1232)$}
\label{sec-finitesizedelta}

In order to check
the influence of the $\Delta(1232)$ resonance also in the finite volume case, we present a calculation of finite size effects in SSE at $\mathcal{O}(\epsilon^3)$ with infrared regularization.

The basic ingredients have already been given in section \ref{deltasec}.
The contribution from the diagram, fig.\,\ref{massdeltagraph},
to the difference $\delta M_N$ between finite and infinite volume becomes \cite{Tim}:
\begin{align}
	\delta^{(3)}_{\text{SSE}} M_N & = 
		\frac{c_A^2 M_0^3}{6 \pi^2 f_\pi^2 M_\Delta^2} 
		\left[\int_0^1 dx +\int_1^\infty dx \right]\sum_{\vec{n}\in\mathbb{Z}^3\backslash\{\vec{0}\}} \nonumber\\
		& \phantom{\times} \left(1-x+\frac{M_\Delta}{M_0}\right) 
		  \bigg[ y(x) K_0 \left(L |\vec{n}| \sqrt{y(x)} \right) \nonumber\\ 
		& - \frac{\sqrt{y(x)}}{L|\vec{n}|} K_1 \left( L |\vec{n}| \sqrt{y(x)} \right) 
			\bigg]
	\label{eq-finitevoldelta}
	\end{align}
where $y(x) \equiv M_0^2(x-1)x + M_\Delta^2 x + m_\pi^2(1-x) $, for details we refer to \cite{Tim}. The finite volume mass-shift at order $\epsilon^3$ is then given as:
\begin{align}
	\delta^{\text{SSE}} M_N(m_\pi,L) &= \delta^{(3)}M_N(m_\pi,L) \nonumber\\ & +\delta_{\text{SSE}}^{(3)} M_N(m_\pi,L)
	\label{rel.SSE}
	\end{align}

\noindent
It has to be added to the infinite volume result of eq.\,(\ref{massaeps3}).
Here we do not perform any truncation in the expansion in powers of $1/M_0$.
Eq.\,(\ref{rel.SSE}) is used with the parameters from ``fit delta II'' to predict the finite volume effects of the nucleon mass with explicit $\Delta(1232)$.

\begin{figure}[tb]
	\centering
	\includegraphics[width=.7\columnwidth,clip=1,trim=10 0 0 -4]{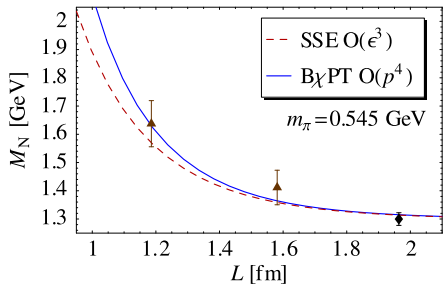} \hfill
	\caption{The box length L dependence of the nucleon mass in relativistic SSE $\mathcal{O}(\epsilon^3)$
	 compared to the B$\chi$PT $\mathcal{O}(p^4)$ result and lattice data at fixed pion mass $m_\pi=545\units{MeV}$.}
	\label{rel545}
\end{figure}

\begin{figure}[tb]
	\centering
	\includegraphics[width=.7\columnwidth,clip=1,trim=12 0 0 0]{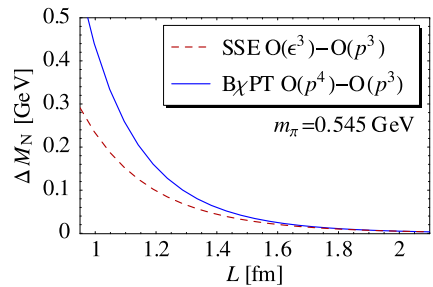}
	\caption{Comparison of $\delta_\text{SSE}^{(3)} M_N(m_\pi,L)$ in eq.\,(\ref{eq-finitevoldelta}) and $\delta^{(4)} M_N(m_\pi,L)$ in eq.\,(\ref{eq-deltamvolp4}).}
	\label{tadp545}
\end{figure}

In fig.\,\ref{rel545} we show a comparison between SSE at $\mathcal{O}(\epsilon^3)$ and
B$\chi$PT at $\mathcal{O}(p^4)$ with the parameters listed in table~\ref{tab-paramsb}.
Both curves are compatible with lattice data. As in the infinite volume case, at the present level of accuracy, the inclusion of explicit $\Delta$ degrees of freedom is not essential to reproduce the trend shown by the lattice data, in the sense that effects of the $\Delta$ can be absorbed in low-energy constants at order $p^4$. In order to quantify the contribution of the $\Delta(1232)$ to the tadpole diagram in B$\chi$PT, we compare $\delta_\text{SSE}^{(3)} M_N$ in eq.\,(\ref{eq-finitevoldelta}) and $\delta^{(4)} M_N$ in eq.\,(\ref{eq-deltamvolp4}) for $m_\pi=545\units{MeV}$, see fig.\,\ref{tadp545}. The overall result gives additional support to our arguments in section \ref{sec-empconstr}, emphasizing the delta-dominance in the low-energy constants entering the tadpole at $O(p^4)$. 

In ref.\,\cite{Beane2} formulae for the finite size effects of the nucleon mass have been presented in the framework of {\it non-relativistic} chiral effective field theory with explicit $\Delta(1232)$ degrees of freedom. We have checked that for pion masses larger than $300\,{\rm MeV}$ and $L <1.5\,{\rm fm}$ the corresponding mass difference between finite and infinite volume lies systematically below the covariant SSE result of eq.\,(\ref{eq-finitevoldelta}), suggesting that recoil corrections could play an important role in this r\'egime \cite{Tim}.

\section{Synthesis} 
\label{sec-voldep2}



\begin{figure}[tb]
	\centering
	\includegraphics[width=\columnwidth]{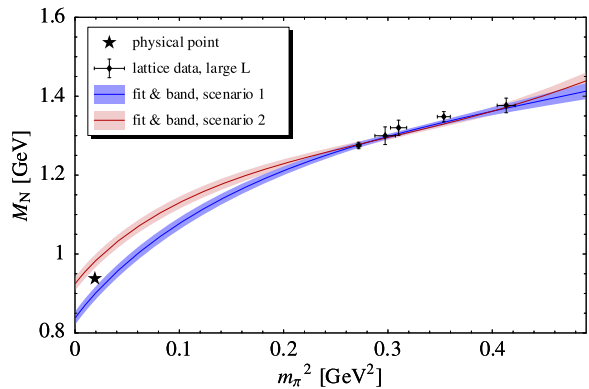}
	\caption{Chiral extrapolation of finite volume lattice data down to the physical pion mass, based on the ${\cal O}(p^4)$ expression in eq.\,(\ref{massp4}). The relevant parameters have been fixed as in table~\ref{tab-synth}. Shown are the resulting $68\%$ local statistical error bands for the nucleon mass extrapolated to infinite volume. The band width can be directly interpreted as the error of a prediction at each given pion mass, see appendix \ref{sec-singlepar}. Only those lattice data points corresponding to large volumes with $L > 2$ fm are displayed. }
	\label{fig-synth}
\end{figure}

With two independent variables at hand, the pion mass $m_\pi$ and the lattice size $L$, one can make use of an enlarged data base to extrapolate down to small quark masses. Figure \ref{fig-synth} shows the results of such a study in the framework without explicit $\Delta(1232)$. Now the physical point is {\it not} included as a constraint. Instead, the low-energy constants $c_i$ have been fixed as input. Two scenarios have been chosen for comparison, with two different sets of $c_i$, as compiled in table  \ref{tab-synth}.
In both cases the finite-volume data with $L < 2$ fm have been included in the fit procedure, in addition to the large-$L$ data shown in fig.\,\ref{fig-synth}. The spin-isospin averaged S- and P-wave $\pi N$ threshold amplitudes are evaluated in parallel, using the rule that $c_3$ is saturated by $\Delta(1232)$ contributions, as elaborated in section 4. The set denoted ``scenario 2'' evidently meets the constraints imposed by pion-nucleon scattering data quite well. The fits show that $M_0$ and $e_1^{(4)}$ are well-determined, and the narrow uncertainty band permits making precise statements at the physical pion mass. 
Note that an exact knowledge of the $c_i$ is crucial to achieve this result. Therefore, in order to release those parameters and keep precision, several observables involving those low-energy constants should be {\it simultaneously} analyzed -- including finite size effects. Such a combined approach promises to be the winning strategy in performing reliable chiral extrapolations.

\newcommand{\cpn}{(\ensuremath{c_3 \rightarrow (4/3)c_3})}
\begin{table*}[tb]
	\caption{Input parameters and resulting output for the nucleon mass extrapolation based on finite volume lattice data. Also listed are the thereshold pion-nucleon amplitudes $T^+$ and $P_1^+$. Consistency with empirical values for these observables is achieved by substituting $c_3 \rightarrow (4/3) c_3$ in their ${\cal O}(p^3)$ HB$\chi$PT expressions, as discussed in section \ref{sec-empconstr}.}
	\label{tab-synth}
\begin{ruledtabular}
\begin{tabular}{ll||ccc} 
	&                                                       &  scenario 1 & scenario 2 &          \\
	\hline 
	$e_1^{(4)}(1 \units{GeV})$ & $(\mathrm{GeV^{-3}})$      &  0.38 $\pm$ 0.04        &  0.62 $\pm$ 0.04      &  fitted     \\
	$M_0$ & $(\mathrm{GeV})$                                &  0.837 $\pm$ 0.017      &  0.924 $\pm$ 0.017    &  fitted     \\
	\hline                                                
	$c_1$ & $(\mathrm{GeV^{-1}})$                           &  $-1.0$              &  $-0.95$           &  fixed      \\
	$c_2$ & $(\mathrm{GeV^{-1}})$                           &  3.2               &  3.0             &  fixed      \\
	$c_3$ & $(\mathrm{GeV^{-1}})$                           &  $-3.4$              &  $-3.65$           &  fixed      \\
	$g_A$ &                                                 &  1.2               &  1.2             &  fixed      \\
	$f_\pi$ & $(\mathrm{MeV})$                              &  90             &  90           &  fixed      \\
	\hline                                                
	$\sigma_N$ & $(\mathrm{MeV})$                           &  55.42 $\pm$ 0.13  &  49.73 $\pm$ 0.14&  \\
	$M_N(m_\pi^\text{phys})$ & $(\mathrm{GeV})$             &  0.899 $\pm$ 0.017      &  0.981 $\pm$ 0.017    &  \\
	\hline                                                
	$T^+$ \cpn & $(\mathrm{GeV^{-1}})$                      &  2.984 $\pm$ 0.021      &  0.101 $\pm$ 0.017    & \\
	$P_1^+$ \cpn & $(\mathrm{GeV^{-3}})$                    &  949.4 $\pm$ 3.0        &  1010.5 $\pm$ 2.3     & \\
	\end{tabular} \par 
	\end{ruledtabular}
\end{table*}


\section{Conclusions}
\label{sec-conclusions}

In this work chiral extrapolations of lattice QCD results for the nucleon mass have been extended in several directions, with the following conclusions to be drawn:

i) A detailed error analysis demonstrates the statistical significance and stability of these extrapolations for pion masses below 0.6 GeV.

ii) For the results to be physically meaningful, it is important to verify that the low-energy constants of the chiral pion-nucleon effective Lagrangian which control this extrapolation are consistent with those extracted from $\pi N$ and $NN$ observables. This is indeed demonstrated to be the case, provided that the $\Delta(1232)$ dominance of the P-wave $\pi N$ parameter $c_3$ is properly accounted for.

iii) Unlike the situation with spin observables such as the nucleon axial coupling constant $g_A$, at the present level of accuracy, it is not crucial to introduce the $\Delta(1232)$ as an explicit, propagating degree of freedom when dealing with the mass of the nucleon. The off-shell propagation length of the $\Delta$ in the $\pi \Delta$ loop correction to the nucleon mass is sufficiently short that its effects can be absorbed in low-energy constants, with the caution exercised as stated in ii).

iv) A highly useful additional source of information from lattice QCD results, apart from their quark mass (or $m_\pi$) dependence, is their variation with the finite lattice size $L$. The observed systematics of this variation supports the underlying chiral dynamics framework, in terms of pion-nucleon degrees of freedom, for the extrapolation to small quark masses. This systematics also reduces ambiguities in constraining input low-energy constants. In particular, the P-wave $\pi N$ parameter $c_3$ preferred by the chiral extrapolation of the nucleon mass turns out to be compatible with the one deduced from $NN$ scattering phases shifts. The larger value found from fits based on ${\cal O}(p^3)$ Heavy Baryon $\chi$PT to threshold pion-nucleon scattering data can be explained in terms of pion-nucleon-delta dynamics in the P-wave channel.

In summary, chiral effective field theory extrapolations of the nucleon mass from lattice QCD to the physically relevant region of small quark masses and further on to the chiral limit are beginning to reach a high degree of consistency with other independent low-energy observables. In order to further improve the accuracy of such investigations, it is an important task for the future (apart from expanding the lattice QCD data base to smaller pion masses) to perform {\it simultaneous} systematic extrapolations of several observables using a single consistent set of low-energy constants in the underlying effective Lagrangian. 
 
\acknowledgments

This work has been supported in part by BMBF.

\appendix
\section{Details of error analysis} 
\label{sec-detailserror}

\subsection{Confidence regions and global error band}

\newcommand{\CL}{\mathrm{CL}}

The information provided by the $N$ lattice data points is of the general form \par
\vspace{5pt}
\begin{centering}
	``$M_N$ \ at \ $m_\pi = x_j \pm \Delta x_j$ \ has \ values \ $y_j \pm \Delta y_j$'' \qquad for \ $j=1,...,N$~. \par
	\end{centering}
\vspace{5pt}
For simplicity, let us first ignore the errors $\Delta x_j$.
We assume the errors in the $y_j$ to be uncorrelated and distributed normally, one
standard deviation being $\Delta y_j$.
We denote our fit function $f(x;p,u)$, with parameters collected in $p=(p_1,...,p_n)$ and
$u=(u_1,...,u_m)$. The parameters $p$ are those which we will estimate from 
lattice data. For the other parameters $u$, most prominently $g_A$ and $f_\pi$,
we take estimates from literature.
During the statistical analysis of uncertainties in $p$, we keep $u$ fixed.
Therefore, we will omit $u$ in our notation. \par
Having agreed on a confidence level $\CL \in (0,1)$,
we are looking for a {\em confidence region} $R_\CL$ fulfilling 
$ P \left[p^\text{true} \in R_\CL \right] = \CL$~,
i.e. the probability to find the true parameters $p^\text{true}$ in the
confidence region is $\CL$. Throughout this work we choose $\CL=68\%$ which
corresponds to one standard deviation for Gaussian distributed errors.

As described in \cite{src-stat,src-minuit,Bernhard}, the $\chi^2$ method offers
a way to construct such regions in good approximation.
Using $\chi^2(y|p) \equiv \sum_{j=1}^N \left(y_j - f(x_j;p) \right)^2 / {\Delta y_j}^2$
we choose
\begin{equation}
	R_\CL :=
	\left\{ p :\  \chi^2(y|p)\ \leq\ \chi^2_\text{opt}(y) + \chi^2_{\CL,n}\ \right\}~,
	\nonumber
	\label{eq-confregdef}
	\end{equation}
where $\chi^2_\text{opt}(y) \equiv \chi^2(y|p^\text{est}(y))$
is the global minimum of $\chi^2(y|p)$
with respect to $p$ occurring at $p^\text{est}(y)$.
The increment $\chi^2_{\CL,n}$ is a fixed number depending on the confidence level $\CL$
and the number of parameters $n$. It is calculated by solving
\begin{equation}
	W^{(n)}_{\chi^2} (\chi^2_{\CL,n}) = \CL~,
	\nonumber
	\label{eq-threshold}
	\end{equation}
where $W^{(n)}_{\chi^2} (t)$ is the cumulative $\chi^2$ distribution for $n$ degrees of freedom.
\par

\newcommand{\mmin}{\ensuremath{\text{min}}}
\newcommand{\mmax}{\ensuremath{\text{max}}}

We have checked that, for our purposes, $\chi^2$ is approximately quadratic in $p$ around the minimum, i.e.
$\chi^2(p^\text{est}+\delta p) \approx  \chi^2(p^\text{est}) +  \frac{1}{2} {\delta p}_k\, A_{k,l}\, {\delta p}_l$ .
For each pion mass $x$, we determine the minimal and maximal values of the fit function, $f_\mmin(x)$ and $f_\mmax(x)$, which can be found in the confidence region
$R_\CL = \{ p^\text{est}+\delta p: \frac{1}{2}\, \delta p_k\, A_{k,l}\, \delta p_l \leq \chi^2_{\CL,n} \}$ .
Shading the region between $f_\mmin(x)$ and $f_\mmax(x)$ results in a global ``statistical error band''
such as shown in fig.\,\ref{fig-band}. It sweeps over all interpolating functions
allowed by statistics within the confidence level. \par

The error matrix $E^{\CL,n}$ and the gradient $v(x)$ of the interpolating function $f(x)$ are defined as
\begin{equation}
	E^{\CL,n} \equiv 2\, \chi^2_{\CL,n}\, A^{-1}\ , \qquad 
	v_k(x) \equiv \left.\frac{\partial f(x;p)}{\partial p_k}\right\vert_{p=p^\text{est}}\ .
	\nonumber
\end{equation}
Denoting $\Delta f(x) \equiv \sqrt{v(x)^\top\, E^{\CL,n}\, v(x)}$, the global error band is approximately given by
$f_{\mmax,\mmin}(x) = f(x;p^\text{est}) \pm \Delta f(x)$~.

\subsection{{\tt Mathematica} Code Sample}

A quick way to obtain a statistical error band is provided by the following {\tt Mathematica 5.2} code snippet.
The function {\tt ErrorBand[data,model,vars,params,npar,CL]} returns 
a list containing expressions for $f_\mmin(x)$ and $f_\mmax(x)$. The first four arguments 
{\tt data}, {\tt model}, {\tt vars} and {\tt params} correspond to those in the
{\tt Mathematica} function {\tt NonlinearRegress}: {\tt model} contains $f(x;p)$,
{\tt vars} lists the independent variable(s) $x$, and {\tt params} specifies the parameters 
$p$, optionally together with initial values for the search. Here {\tt data} is a list of tuples
of the form  \{$x_j$,$y_j$,$\Delta y_j$\}, i.e. it has an additional column specifying
the one-standard-deviation errors $\Delta y_j$. {\tt npar} is typically set to {\tt Length[params]}
so as to generate the global error band, but it must be set to $1$ when a local error band
as discussed in section \ref{sec-singlepar} is desired. The confidence level {\tt CL} defaults
to $68\%$, corresponding to one standard deviation. If options are given, they are passed to
{\tt NonlinearRegress}.

Please note that {\tt ErrorBand[...]} uses an error matrix which is only a good approximation when the model $f(x;p)$
is sufficiently linear in the parameters $p$ and the fit is good, cf. eq.\,(4.62) in ref.\,\cite{Bernhard}.
A direct calculation of the error matrix, as done, for example, by {\tt MINUIT}, does not suffer from this limitation. 

\begin{widetext}
{\small \begin{verbatim}
<< Statistics`NonlinearFit`;
<< Statistics`NormalDistribution`;

ChiSqrThresh[n_, CL_] :=
  x /. Solve[CDF[ChiSquareDistribution[n], x] == CL, x][[1]];

ErrorBand[data_,model_,vars_,params_,npar_,CL_:0.68269,options___]:=
  Module[{reg,emat,grad,thr,err,p},
    thr = ChiSqrThresh[npar,CL];
    reg = NonlinearRegress[Drop[data,{},{-1}],model,vars,params,Weights->1/(Last/@data)^2,
            RegressionReport->{BestFit,BestFitParameters,EstimatedVariance,AsymptoticCovarianceMatrix},
            options];
    (* extract approximation to error matrix *)
    emat = thr * (AsymptoticCovarianceMatrix/.reg)[[1]] / (EstimatedVariance /. reg);
    (* gradient of fit function with respect to the parameters *)
    p = (Flatten/@(List/@params))[[All,1]];
    grad = (D[model,#]&/@p) /. (BestFitParameters/.reg);
    (* calculate parabolic error *)
    err=Sqrt[grad.emat.grad];
    {(BestFit/.reg)-err,(BestFit/.reg)+err}
  ];
\end{verbatim}}
\end{widetext}

\subsection{Systematic envelope}

Up to here we have neglected uncertainties concerning the fixed parameters $u$, such as $g_A$ and $f_\pi$.
While we have a good guess about acceptable ranges for their values, we have no knowledge about possible correlations of these error estimates. Therefore we scan the whole
range of acceptable values of $u$ on a grid, perform the complete statistical error analysis
for each choice of $u$ and pick the extreme values of $f_\mmin(x)$ and $f_\mmax(x)$.
We call the resulting region the ``systematic envelope'' in our plots. 

\subsection{Single-parameter errors and local error band}
\label{sec-singlepar}

Confidence intervals for any parameter dependent quantity $q(p)$ satisfying $P\left[ q_\mmin \leq q(p^\text{est}) \leq q_\mmax \right] = \CL$ can be generated in good approximation using the $\chi^2$ method:
\begin{equation}
	[q_\mmin,q_\mmax] = \left\{\ q(p) \ : \ \chi^2(y|p)\ \leq\ \chi^2_\text{opt}(y) + \chi^2_{\CL,1}\ \right\}~,
	\nonumber
	\label{eq-1paramthresh}
	\end{equation}
where now the single-parameter increment $\chi^2_{\CL,1}$ has to be employed.
For a confidence level $\mathrm{CL}=68\%$ corresponding to a one standard deviation error, $\chi^2_{\CL,1}=1$.

A simple case of such a parameter dependent quantity $q(p)$ is a specific parameter $p_k$ itself.
With the quadratic approximation of $\chi^2$ from above, the single-parameter errors appear on the diagonal of $E^{\CL,1}$: $(\Delta p_k)^2 = E^{\CL,1}_{k,k}$. Note that the multi-parameter confidence region is not enclosed in the box of single-parameter error bounds \cite{src-minuit}.

As another important case, consider a {\it local} error band, formed by plotting the single-parameter confidence interval for $f(x;p)$ at every $x$. It is narrower than the corresponding global band and needs to be interpreted differently. Its construction is identical to that of the global band, except that now the error threshold is $\chi^2_{\CL,1}$, and the error matrix is $E^{\CL,1}$. We opt to show such a band in fig.\,\ref{fig-synth}, where we want to read off the error of our nucleon mass prediction directly.

\begin{table*}[tb]
	\caption{Two-flavor lattice data for the nucleon mass $M_N$ taken from \cite{QCDSF,src-Orth}
	and selected according to the criteria described in appendix \ref{sec-latticedat}. The index numbers in the first column comply with those in the compilation \cite{QCDSF}.}
	\label{tab-latticedat}
	\begin{ruledtabular}
	\begin{tabular}{rl|llrr|rr|c} 
	no. & collaboration & 
	$\beta$ &
	$\kappa$ &
	$a\ [\mathrm{fm}]$  & 
	$L\ [\mathrm{fm}]$ & 
	$m_\pi\ [\mathrm{GeV}]$ & 
	$M_N\ [\mathrm{GeV}]$ &
	large $L$ \\
	\hline 
	19 & CP-PACS & 2.1 & 0.1382 & 0.111 & 2.68 & 0.5214(21) & 1.2751(82) & $\times$ \\
	\hline
	41 & JLQCD & 5.2 & 0.1355 & 0.098 & 1.96 & \underline{0.5453(91)} & 1.300(23) & $\times$ \\
	36 & JLQCD & 5.2 & 0.1355 & 0.099 & 1.58 & 0.560(16) & 1.412(62) &  \\
	31 & JLQCD & 5.2 & 0.1355 & 0.099 & 1.19 & 0.655(32) & 1.637(82) &  \\
	\hline
	8 & QCDSF & 5.25 & 0.13575 & 0.092 & 2.21 & 0.5570(70) & 1.320(20) & $\times$ \\
	\hline
	23 & CP-PACS & 2.2 & 0.1368 & 0.092 & 2.22 & 0.5946(53) & 1.348(13) & $\times$ \\
	\hline
	56 & \cite{src-Orth} & 5.6 & 0.1575 & 0.085 & 2.04 & \underline{0.6429(68)} & 1.377(19) & $\times$ \\
	55 & \cite{src-Orth} & 5.6 & 0.1575 & 0.084 & 1.34 & 0.660(12) & 1.471(29) &  \\
	54 & \cite{src-Orth} & 5.6 & 0.1575 & $\sim$0.085 & 1.19 & 0.709(11) & 1.672(38) &  \\
	53 & \cite{src-Orth} & 5.6 & 0.1575 & $\sim$0.085 & 1.02 & 0.832(22) & 1.900(39) &  \\
	\end{tabular} 
	\end{ruledtabular}
\end{table*}

\subsection{Errors in the pion mass}
\label{sec-errpionmass}

The errors $\Delta x_j$ in the pion mass can be treated by setting up $\chi^2$ as
\begin{equation}
	\chi^2(p,\delta x) = \sum_{j=1}^N \left( 
		\left[ \frac{f(x_j+\delta x_j;p) - y_j}{\Delta y_j} \right]^2
		+ \left[ \frac{\delta x_j}{\Delta x_j} \right]^2
		\right)~.
	\nonumber
	\end{equation}
The minimization and error treatment must now be performed on $n+N$ parameters $p$,~$\delta x$.
Since we are not interested in the joint confidence region of $p$ and $\delta x$,
but only in the confidence region of $p$, it makes sense to continue using the
threshold value $\chi^2_{\CL,n}$ and to keep $\chi^2$ minimized with respect to the $\delta x$
at all times \cite{Bernhard}. \par
We can avoid too many parameters by making an approximation. Assuming
that $f(x_j+\delta x_j|p)$ is approximately linear in $\delta x_j$
for $\delta x_j \lesssim \Delta x_j$ and minimizing $\chi^2$ with respect to the $\delta x_j$ yields \cite{Bernhard}
\begin{multline*} 
	\chi^2_\text{eff}(p) =  \\
	 \sum_{j=1}^N \frac{ \left[f(x_j;p) - y_j \right]^2}
	{ \left[\Delta y_j \right]^2 
	  + \left[ f(x_j+\Delta x_j /2\, |\,p) - f(x_j-\Delta x_j /2\, |\,p) \right]^2 }
	\end{multline*}
Within our approximation, $\chi^2_\text{eff}$ has the same minimum as the original, full $\chi^2$. For 
hypothesis testing, the number of degrees of freedom remains $2N-(N+n)=N-n$.
For confidence regions for $p$, the $\chi^2$ increment is $\chi^2_{\CL,n}$
as before.


\section{Lattice Data}
\label{sec-latticedat}

The lattice data used for our calculations was taken from refs.\,\cite{QCDSF,src-Orth} and is listed in table \ref{tab-latticedat}.
The conversion into physical units was performed setting the Sommer scale $r_0 = 0.5 \units{fm}$ \cite{QCDSF}. 
Related systematic uncertainties need to be estimated by further lattice studies, and are not discussed throughout this work.
The table is divided by horizontal lines into ``volume groups'' characterized by the same value of simulation parameters $\beta$ and $\kappa$. The data have been selected according to the following cuts: $L > 1\units{fm}$, $a < 0.15\units{fm}$. In addition, the point of largest simulation volume in each group must fulfill $m_\pi < 0.65\units{GeV}$ and $m_\pi L > 5$. 
The lattice data from \cite{src-Orth} have been generated with a standard, unimproved Wilson fermion action.
Even though, lattice artefacts are shown \cite{src-Orth} to be small for the points we select here, in virtue of the fine lattice spacing. 
For points $53$ and $54$, no calculation has been performed for the Sommer radius $r_0/a$, which is needed to determine the lattice spacing $a$. In accordance with ref.\,\cite{src-Orth}, $a$ for these two points is copied from point 56.

\rule{0cm}{0pt} \\
\rule{0cm}{0pt} \\
\rule{0cm}{0pt} \\
\rule{0cm}{0pt} 


\end{document}